\documentclass[12pt]{article}
\usepackage{graphicx}
\usepackage{amssymb,amsmath,amsfonts,palatino,amsthm}
\usepackage{amssymb}
\usepackage{epstopdf}
\usepackage{slashed} 
\newcommand{\f}{\begin{equation}}
\newcommand{\ff}{\end{equation}}

\begin{document}

\title{Reversing the Irreversible: from limit cycles to emergent time symmetry \\}
\author{Marina Cort\^{e}s${}^{1,2,3}$ and Lee Smolin${}^{1}$
\\
\\
Perimeter Institute for Theoretical Physics${}^{1}$\\
31 Caroline Street North, Waterloo, Ontario N2J 2Y5, Canada
\\
\\
Institute for Astronomy, University of Edinburgh ${}^{2}$\\
Blackford Hill, Edinburgh EH9 3HJ, United Kingdom
\\
\\
Instituto de Astronomia e Ci\^{e}ncias do Espa\c{c}o${}^{3}$\\
Faculdade de Ci\^encias, Edif\'{i}cio C8, Campo Grande, 1769-016 Lisboa, Portugal}
\date{\today}
\maketitle

\begin{abstract}
In 1979 Penrose hypothesized that the arrows of time are explained by the hypothesis that the fundamental laws are time irreversible\cite{Penrose1979}. That is, our reversible laws, such as the standard model and general relativity are \textit{effective}, and emerge from an underlying fundamental theory which is time irreversible. 
In \cite{ECS1,ECS2,ECS3} we put forward a research program aiming at realizing just this. The aim is to find a fundamental description of physics above the planck scale, based on irreversible laws, from which will emerge  the apparently reversible dynamics we observe on intermediate scales.

Here we continue that program and note that a class of discrete dynamical systems are known to exhibit this very property: they have an underlying discrete irreversible evolution, but in the long term exhibit the properties of a time reversible system, in the form of limit cycles. 
We connect this 
to our original model proposal in \cite{ECS1}, and show that the behaviours obtained there can be explained in terms of the same phenomenon: the attraction of the system to a basin of limit cycles, where the dynamics appears to be time reversible. Further than that, we show that our original models exhibit the very same feature: the emergence of quasi-particle excitations obtained in the earlier work in the space-time description is an expression of the system's convergence to limit cycles when seen in the causal set description. 

\end{abstract}

\newpage

\tableofcontents


\section{Introduction}

One of the most mysterious questions facing cosmology is to account for the several arrows of time.  If we accept the usual assumption that the fundamental laws are time reversible, the ubiquity of the arrows of time can only be explained to be a consequence of extremely special and improbable initial conditions.  This was Boltzmann's explanation for the paradox of how the time asymmetric second law of thermodynamics is to be derived, as emergent from the statistical treatment of time symmetric microscopic laws.  The problem with this is that one then has to account for the vastly improbable initial conditions which must be imposed at the beginning of the universe.  

Back in 1979, Roger Penrose proposed a different explanation for the arrows of time in \cite{Penrose1979}, which we take up in this note.  We posit, with Penrose, that there is another layer of fundamental law below general relativity and quantum theory which is {\it time asymmetric.}   However after a transient period, time reversible behaviour emerges naturally.  Thus, after a certain time, the time asymmetric fundamental dynamics can be approximated by a time symmetric effective dynamics, supplemented by time asymmetric initial conditions.  

Thus, when we observe a highly time asymmetric history, in which, for example, electromagnetic waves excited by point like sources propagate outwards towards the future, but never towards the past, we are used to believing that the time reversed history, in which waves excited by local sources propagate to the past, are also possible, as they are solutions
to time symmetric laws.  We don't see them because they are forbidden by highly improbable, time asymmetric, initial conditions.
In the alternative viewpoint we endorse here, only the histories with outgoing propagation to the future are  solutions to the  truly fundamental, {\it time asymmetric}, laws.  Hence the time asymmetric universe which is often explained by a combination of time symmetric laws and time asymmetric initial conditions is, we envision, to be more simply explained by time asymmetric laws.  This requires that for long times, some of the solutions to the time symmetric theory are also to a good approximation to a more fundamental time asymmetric theory.  But the time reversals of those solutions are not solutions to the fundamental theory.  

Here we support Penrose's hypothesis, by first showing that this emergence of apparently time symmetric dynamics from time asymmetric dynamics is ubiquitous in a large class of dynamical systems.
These are systems that have the following properties: 

\begin{enumerate}

\item{}There are a large, but finite number of discrete states $S= \{ I,J,K,L, \ldots \}.$  

\item{}There is a discrete and deterministic evolution rule. Thus each state, $I$, has a unique successor:
\f 
I \rightarrow J
\ff
\item{}There is no rule constraining the number of precedents a state $I$ may have.  In particular, there may be states with several precedents, as well as ``garden of eden states" with no precedent.  

\end{enumerate}

Generic dynamical rules of this kind are irreversible.  But it is well known that such systems evolve to limit 
cycles\cite{AW}.  This is for the simple reason that the evolution rule must give each state a unique successor picked from a finite set of states.  So the evolution must return within a finite number of steps to a previous state, after which, by the rule of unique succession, it cycles.  

The evolution around these limit cycles, once entered, have unique precedents as well as unique successors, and hence can be described by reversible dynamical laws, supplemented by time asymmetric initial conditions, which impose the fact that due to the unique successor rule, the original dynamics allows each cycle to be traveled in only one direction. 

This achieves three ends.  First we give a large class of examples of systems that start off irreversibly and evolve to states which may be described by reversible evolution rules supplemented by irreversible initial conditions.  Second, we show that this behaviour is ubiquitous in discrete systems.  If the universe is ultimately described by a discrete dynamical system then this is a natural and simple explanation for the existence of an arrow of time.

The hypothesis that early universe cosmology was driven by an irreversible dynamics, which gave rise to an emergent reversible dynamics, is explored also in \cite{ECS1,ECS2,ECS3}.  
There we presented a class of modified causal set models, which we called 
{\it energetic causal set models}\footnote{Other versions of causal set models were introduced earlier 
by\cite{cs,cohl}.}. 
These have an irreversible dynamics.  In the first of these
papers \cite{ECS1} we studied numerically a $1+1$ dimensional  energetic causal set model.  We found that the model exhibits two phases.  The initial phase is chaotic and clearly irreversible.  It is followed by a transition to a second, more ordered phase, which appears to be dominated by a reversible regime.  

In section 4 we show that the long term behaviour of this model is dominated by one of a small number of limit cycles.  This explains the emergence of quasi reversible dynamics in these models. This is the third and main conclusion of this papre.

Two comments before proceeding. First, the properties of discrete dynamical systems and, in particular, the ubiquity of limit cycles is well known, so nothing new is claimed about those.  Second,  this argument supports the claim that time is real rather than emergent in nature, as it makes little sense to posit an emergent but irreversible time.  If time is emergent or reducible to a block universe prospective, so that the future is no different from the past, there cannot emerge an irreversible dynamics.  But if time is fundamental so that the future has a
 very different status from the past, it is natural to postulate an irreversible dynamics.  Thus, the results of this paper contribute to the research program set out in \cite{SURT,TR}.

\section{Discrete dynamical systems and limit cycles}
\label{finiteDDDS}
Consider a general finite state deterministic discrete dynamical system (DDDS for short): this can be described as a finite set of states, 
$S = \{ I,J,K,L, \ldots \}$, with a successor rule, so that each state has a unique successor 
\f
I(t) \rightarrow I(t+1) 
\ff
or 
\f
I \rightarrow J \rightarrow K \rightarrow ...
\ff
here time, $t$, is discrete so $t$ is an integer.

\begin{figure}[t!]
\centering
\includegraphics[width= 0.95\textwidth]{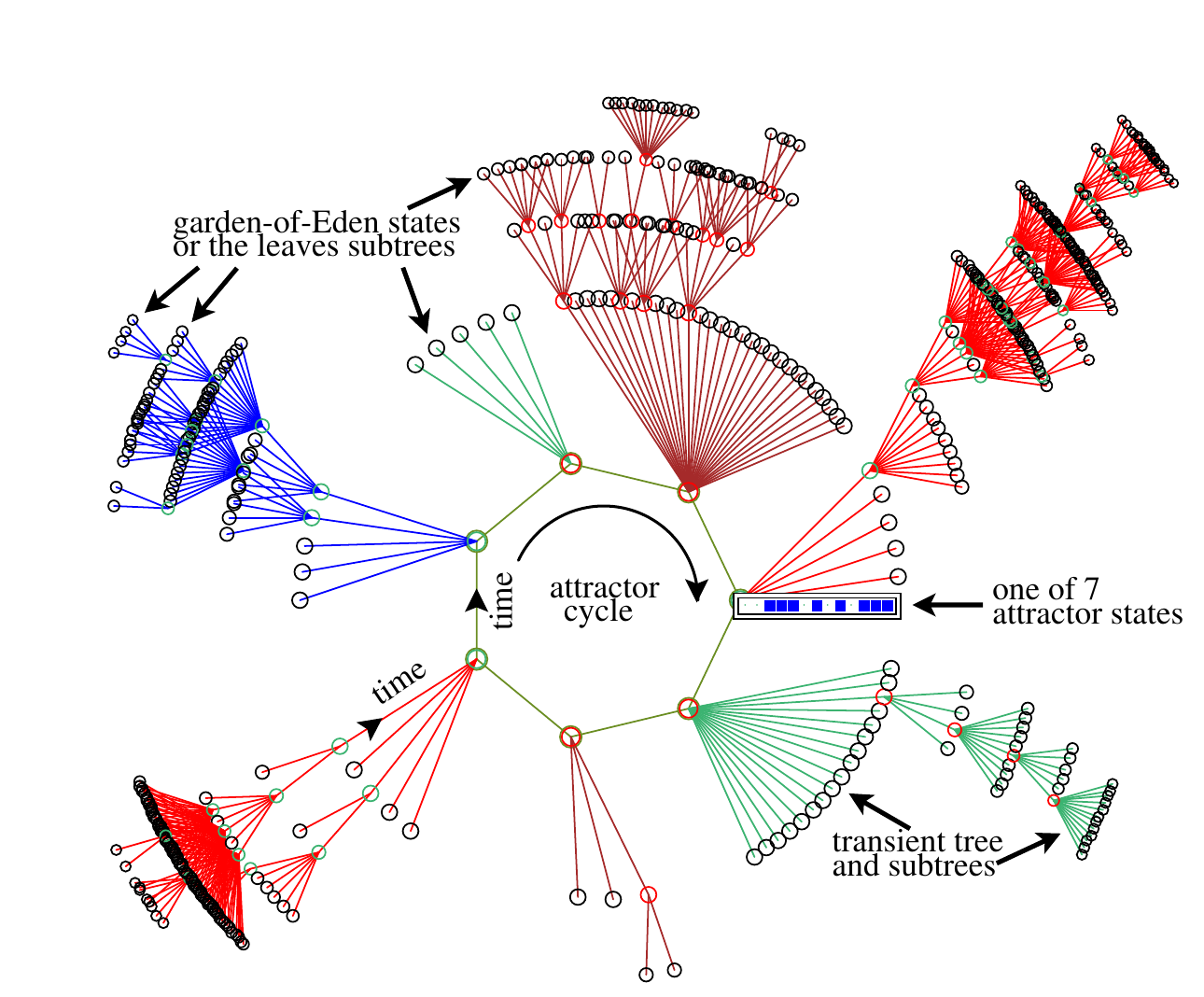}
\caption{The basin of attraction of one limit cycle in a discrete deterministic dynamical system. Image courtesy of Andy Wuensche from \cite{AW}.}
\label{DDS1} 
\end{figure}

A given state can have more than one parent, but each has only one child or successor.  So a general DDS is irreversible.

A limit cycle is when a bunch of states form a closed loop under the successor rule:
\f
Cycle =C:Ê I \rightarrow J \rightarrow K \rightarrow L \rightarrow I 
\ff
Once the system gets into a cycle it will stay there forever.

There can be several ways to enter the cycle,
\f
X \rightarrow K \rightarrow L \rightarrow I\rightarrow J \rightarrow K \rightarrow L 
\ff
etc but once in you are trapped forever.

Now WITHIN THE CYCLE, each state has a unique parent and a unique child.Ê So a reverse cycle exists as a possibility, even if it is not realized:
\f
C^* ;  L \rightarrow K \rightarrow J \rightarrow I \rightarrow L 
\ff
Hence once an observer realizes the system is trapped in a cycle, they could describe the system by a time reversible effective dynamics.  The real dynamics is not reversible and the reverse cycle never happens, but in the effective dynamics, it is as if it could happen.  

Hence, {\it it is natural and generic that a general discrete dynamical system is characterized by a transition from a fundamental irreversible behaviour to an emergent reversible behaviour.}

\begin{figure}[t!]
\centering
\includegraphics[width= 0.95\textwidth]{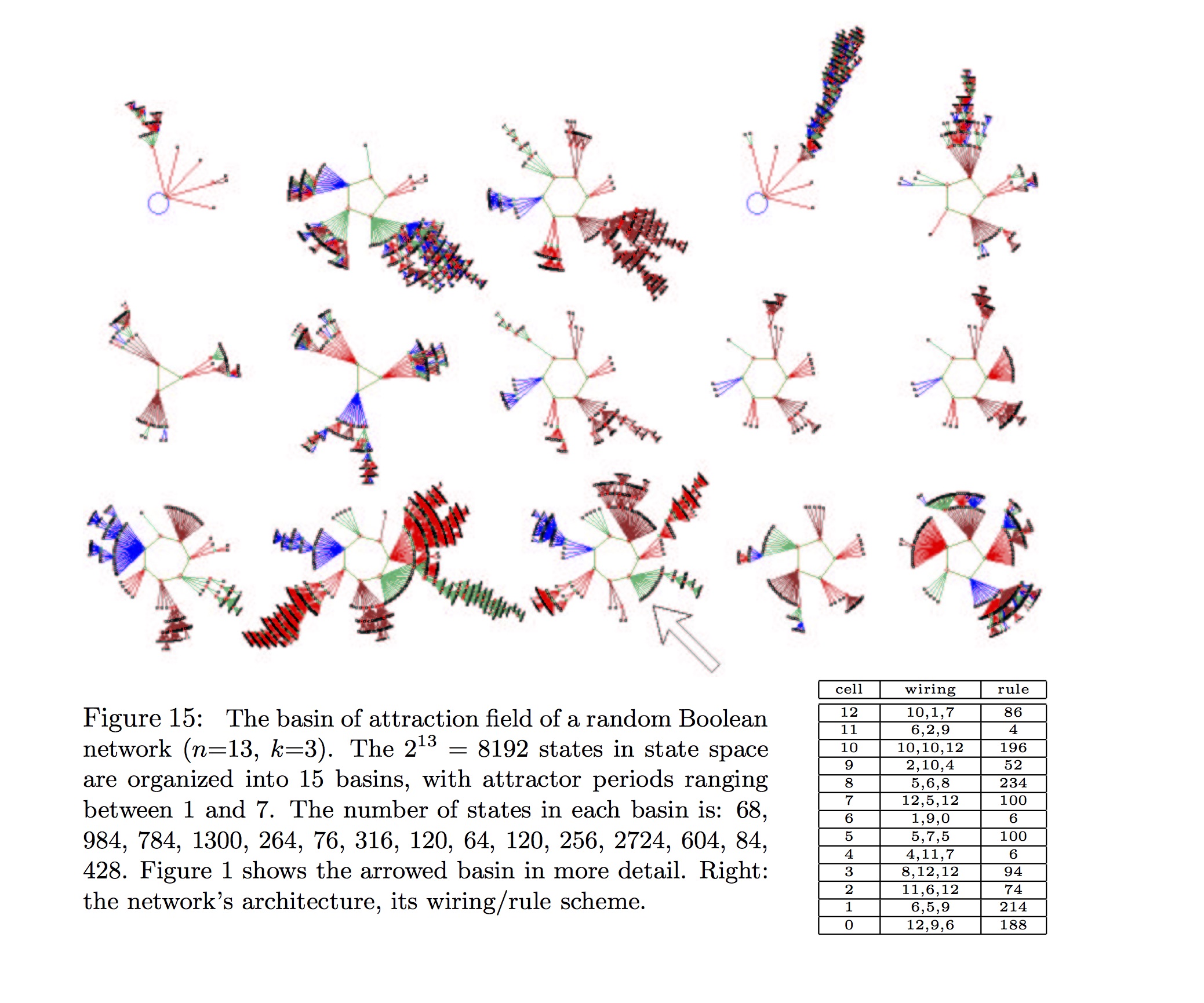}
\caption{The complete state space of a discrete deterministic dynamical system, showing several limit cycles and their basins of attraction.  Image courtesy of Andy Wuensche from \cite{AW}.}
\label{DDS2} 
\end{figure}

Note that within the given $DDS$ the time reversed behaviour never happens. Since each state has a unique successor, if a cycle $C$ is part of a $DDS$, its time reverse $C^*$ is not part of the $DDS$.  However if you chanced upon the system after it has entered the cycle, you would notice that each state has both a unique successor and unique predecessor and you could be fooled into representing the system as having a reversible dynamics.

Now, suppose you came upon a collection of copies of the same $DDS$.  You would be surprised to see that in all cases the systems were going around the cycle $C$, while in no cases were systems going around the time reversed cycle $C^*$.  You would be tempted to speculate that the universe has a "mysterious arrow of time" which (since you would mistakenly believe the underlying dynamics is reversible) can only be explained by "mysterious and highly improbable initial conditions" in which all the members of the collection started off going around the cycle in the same direction.  In fact, the true explanation is that the cycle $C$ is only a limit cycle of a larger irreversible $DDS$.  So the apparent "improbable initial conditions" are actually explained by a deeper irreversible dynamics.  This is what we assert could be true in nature and explain the mysterious arrows of time in cosmology.

Now consider the limit set of S, this is a subset L in S of states that are gotten to after a large number of steps $M>> N$, starting at any state in S.  It is easy to see that L consists of one or more cycles, L= union of cycles, C.  
But this means that for any DDS, S, its limit set, L,  has an effective reversible dynamics.

We can call this kind of behaviour {\it mock-reversible}.  A system is mock-reversible if its evolution has the property that each state in the limit set has unique children and parent states, so that the time reverse
of the dynamics is defined on the limit set.  The reversibility is only {\it mock} because the time reversed histories are  not actual solutions of the dynamics.  Normally they would be eliminated by initial conditions, instead they are eliminated by the dynamics.  


\section{Random discrete dynamical systems}

Consider an ensemble of random discrete dynamical systems each with $N$ states, where each state $J$ (except $I$) has an equal probability, $ p = \frac{1}{N-1}$, of being the succeser of state $I$.

It is easy to show that in the case that $N$ is large\footnote{For a detailed study and review of random boolean networks, which are a large class of examples of these systems, 
see\cite{BD-RBN}.},

\begin{itemize}

\item{}The length of a typical limit cycle is $L_C$ is 
\f 
< L_C > \approx \sqrt{\frac{N}{2}} 
\ff 

\end{itemize}

The argument is as follows:

Choice a random starting point, state $I$, which has a unique successor $J$.  The probability that $I$ is the successor of $J$ is $p_{I \rightarrow J \rightarrow I} = \frac{1}{N} $.  If this is not the case, then we have $I \rightarrow J \rightarrow K$ and the probability that either $I$ or $J$ is the successor of $K$ is $ \frac{2}{N} $.  It follows that the probability that the series closes after $M$ steps, to somewhere along it is
\f
p_{closes} = \frac{1}{N} + \frac{2}{N} + \ldots + \frac{M}{N} = \frac{M(M+1)}{2N}
\ff
The appearance of a cycle becomes likely when this approaches one, so that a typical length for a 
transient from an arbitrary starting point to a cycle, plus once around the cycle is $M \approx \sqrt{2N}$.   Now the cycle is equally likely to close anywhere along these $N$ steps, so that
\f
M_C = <L_C>  \approx \sqrt{\frac{N}{2}}
\ff
The typical number of steps from an arbitrary starting point along a transient to a point on a cycle is then also,
\f
<M_T>  \approx \sqrt{\frac{N}{2}}
\ff
But the arbitrary starting point is not the origin of the transient.  That origin, a so-called ``garden of eden state" is likely to be the same distance from the starting point as the cycle, hence the typical length of a transient is:
\f
<L_T>  \approx 2 \sqrt{\frac{N}{2}} =  \sqrt{2N} 
\ff

Note that the system can be represented by a directed graph $\Gamma$ with $N$ links.

Now,  each state has exactly one successor.  This means that on the average each state has a single predecessor. But note that there are several ways this average could be achieved.  For example, if there are $N/2$ garden-of-eden states, then all the other states will have an average of $2$ predecessors.    At the other extreme, if there are no garden of eden states then the graph is a collection of closed loops and there are no transients.  In this case there are roughly 
\f
N_{cycles} \approx \frac{N}{L_C} \approx \sqrt{N}
\ff
cycles,  of average length $L_C = \sqrt{N}$.  

Now, suppose that the number of garden of eden states is much less than $N$.  Then most states have just one predecessor.  In this case, the  $\sqrt{\frac{N}{2}}$ states in a cycle will have $O(1)$ transients approaching it. The basin of attraction of a typical cycle is then of size order $L_B \approx O(\sqrt{N})$.  There are then again of order
\f
N_{cycles} \approx \frac{N}{L_B} \approx \sqrt{N}
\ff
cycles, each with a basin of attraction.

\section{Converging to limit cycles - Energetic Causal Sets}

In this section and for the remainder of the article we focus on the convergence to limit cycles by energetic causal sets. ECS are models proposed by us in \cite{ECS1,ECS2,ECS3} and incorporate fundamental time asymmetry, causality and discreteness to give origin to a discrete irreversible causal model for spacetime.

In numerical studies of a $1+1$ dimensional version\cite{ECS1}, evidence was noted for a transition from an early, disordered phase to a later, ordered, phase.  The earlier phase was characterized by irreversiblity, while the later phase appeared to be characterized by the emergence of reversible phenomena such as the propagation and scattering of quasi-particle excitations.  This appeared also to involve the collapse of the degrees of freedom to a small limit set, shown by the regular trajectories of quasi-particles.  This naturally raises the question of whether the apparent emergence of reversible dynamics is related to the contraction of the state space to limit cycles, as discussed above.

In this section we show that the answer is yes. Above we have discussed deterministic DDS's and their respective attractions to the basin of limit cycles. We will then first study a specialization to a deterministic  case of ECS, and then generalize to the larger set of ECS which have evolution rule which are, typically, stochastic .


One might have expected the deterministic form of the ECS to naturally exhibit limit cycles, since it satisfies all the requisites discussed above. However, for for the stochastic model, the lack of determinism could threaten the regularity of the evolution which is key to the formation of cycles in the basin of attraction.

Despite this we show, quite surprisingly, that even in their stochastic incarnation, ECS show a strong attraction to limit cycles in their typical dynamical regime. The system will be caught in a particular limit cycle, then due to stochasticity occasionally jump out of it, and will be immediately caught by another limit cycle, see Section~\ref{SectionStochECS}.

Once the system reaches the limit cycle regime, the effect of the randomness in the evolution is to occasionally perturb the system from one limit cycle to another\footnote{This effect is illustrated in the "jump graph", section 20.3 in \cite{AW}.}.

Hence we proceed in two steps.
We first study a deterministic specialization of the $1+1$ dimensional ECS model.  We show that it can be mapped to a finite state DDDS (discrete, deterministic, dynamical system), so
that the long term behaviour is indeed determined by limit cycles. After this we study the original stochastic energetic causal set model we introduced in \cite{ECS1} and show that it quickly evolves to a regime of successive limit cycles and stays there. 

\subsection{Generating energetic causal sets}

The (thick) present state of the system consists of $N$ events on a $1+1$ dimensional 
cylinder\footnote{For definitions and details, please see \cite{ECS1}.}, 
${\cal C} = S^1 \times R $, coordinatized by $x^a = (t, x)$ and subject to identifications 
\f
x \sim x + n L
\label{pos}
\ff
where $n$ is an integer describing revolutions or `windings' wrapped around $S^1$. We impose the Minkowski metric.



In energetic causal sets an event is defined by the intersection of two momenta. Each event has therefore two incoming and two outgoing momenta. For simplicity we'll use null momenta and then we 
will find it convenient to employ null coordinates $z^\pm = t \pm x$. Each event is then represented by a pair $z_i^{\pm}=(z_i^+,z_i^-)$ subject to
identifications
\f
(z^+ , z^-) \sim (z^+ + n L , z^- - nL)
\ff


The rule by which new events are generated is a function of the history  of the events in the present state.  What is relevant for this rule is the causal past of each event.  In ECS each event has a set of ancestor events which we call its {\it past} or {\it lineage}. 

The past of an event is summarized by a number -- the average of the $x$ positions of all of its ancestors, $\Sigma (z^\pm_i)$.
As we show below, the pasts are going to determine which events create the present. 
For computational simplicity we only trace either the left or right ancestry, 
with no change to results. The pasts of events in the initial time-like slice $t=0$ is their current $x$. \footnote{The idea of taking the set consisting of the space-time position of an event together with its past, and its influence on the generation of new events, is the key concept in the creation of the original ECS model. It ensures uniqueness of universal events in cosmology, which is the starting point of the framework. Here we show that this model, as proposed initially, already exhibits limit cycles.}



Therefore each event in the thick present, $z^\pm_i \in {\cal C}$ has a past, $\Sigma (z^\pm_i)$. A set of pairs of events in the present, plus their pasts, $\{z^\pm_i,\Sigma (z^\pm_i)\}$,  is called a state, $\Psi$.

\subsubsection{Running the model}
\label{algorithm}

We follow closely the prescription of \cite{ECS1} which can be summarized as 

\begin{itemize}

\item{} {\bf STEP 0}: Initialize.

Pick $N$ initial events $z^\pm_i \in {\cal C}$, 
all relatively space like.  The number of initial events $N$ corresponds to the number of pasts or lineages in the model, and the initial values of the pasts is simply the initial $x^1$.This is the initial state, $\Psi_0$ and it constitutes the present. 

\item{}  {\bf STEP 1}:  Choose the events which will interact:

Next, we choose two events from the set of present events which will interact to create a new event.  To do this, we define a positive fitness function $F(z^+_i, z^-_j )$ on 
all pairs of events in {\cal C}, to be the absolute value of the difference between the pasts of those two events,
\f
F(z^+_i, z^-_j )= | \Sigma (z^+_i) - \Sigma (z^-_j)  |\,.
\ff 

Now pick the pair which minimizes $F$, and call it  $\{z^\pm_{*} , y^\pm_*\}$. This will be the pair which will interact and generate the new event.

An event can only have two children.  So we next have to ascertain which (if any) of the two possible pairs of outgoing momenta have not yet caused a future event, and thus are available. If the most fit pair $\{z^\pm_{*} , y^\pm_*\}$ has both pairs of momenta available, then choose randomly (or deterministically) which pair will interact. This, incidentally, will be one of the two steps in the dynamics of the model which determines if the algorithm is deterministic or stochastic. If only one pair is available, use that one. If no pair of momenta of available choose the next most fit pair.  
 


When an event has both pairs of momenta available for interaction we can choose randomly or deterministically which pair will interact. Once an event has used both momenta, it is no longer part of the present.



\item{}  {\bf STEP 2}:  Create a new event: 

Take $\{z^\pm_{*} , y^\pm_*\}$ and choose a value for the integer $n$ defined in Eq.~\ref{pos}. The choice of $n$ is the second choice which can be made deterministically or randomly and, together with the choice mentioned above in Step 1, alters the overall performance of the model. Then we create a new event by
\f
z^{\pm}_{\rm new} = (z^+_* +nL, y^-_* -nL),
\ff
where we have ordered the events so $z_*^+ < y^-_*$. 




Add the new event  $z^{\pm}_{\rm new}$ to the present set of events, and assign it to the chosen lineage of its ancestors, $z^\pm_{*}$ or $y^\pm_*$. Add the new pair to $\{z^{\pm}_{\rm new},\Sigma(z_{\rm new}^{\pm})\}$ to $\Psi_0$.

If either $z_*^{\pm}$ or $y_*^{\pm}$ have both momenta used erase them from $\Psi_0$.
Make the step $\Psi_0 \rightarrow \Psi_1$ to obtain the new present \footnote{We refer to Section 4 of \cite{ECS1} for full details of the algorithm for simulating energetic causal sets}. Note that, at each step $\Psi_i$ contains both the events in the present, along with the values of their pasts, so $\Psi_i$ is the set of pairs $\left\{z_i^{\pm}, \Sigma(z_i^{\pm})\right\}$ in the thick present. 

\item{} We then iterate STEPS 1 and  2.

\end{itemize}

\subsection{Converging to limit cycles - deterministic energetic causal set model}
\label{SectiondetECS}

The process described above, the most general ECS, is a discrete dynamical system. We now want to investigate whether it exhibits limit cycles, that is, whether the dynamics will converge to an apparent time reversible evolution. 

In Section~\ref{finiteDDDS} we showed that a finite state deterministic DDS converges to limit cycles. Therefore here we will first specialize to the deterministic version of ECS to investigate the same argument. This is simply done by creating a rule for the moves mentioned in Steps 1 and 2 above, the choice of the pair of photons and winding number $n$. We will systematically choose the left hand pair of photons (having investigated that no loss of generality ensues) and set $n=1$.

The next and final step for us to be able to use the argument of Section~\ref{finiteDDDS} is to assert whether the number of possible states of a given model, i.e. the number of different $\Psi_i=\left\{z_i^{\pm},\Sigma(z_i^{\pm})\right\}$ is finite. 
 
The algorithm generates a discrete sequence of states 
\f
\Psi_0 \rightarrow \Psi_1 \rightarrow \Psi_2 \rightarrow \ldots
\ff

\begin{figure}[t!]
\centering
\includegraphics[width= 0.95\textwidth]{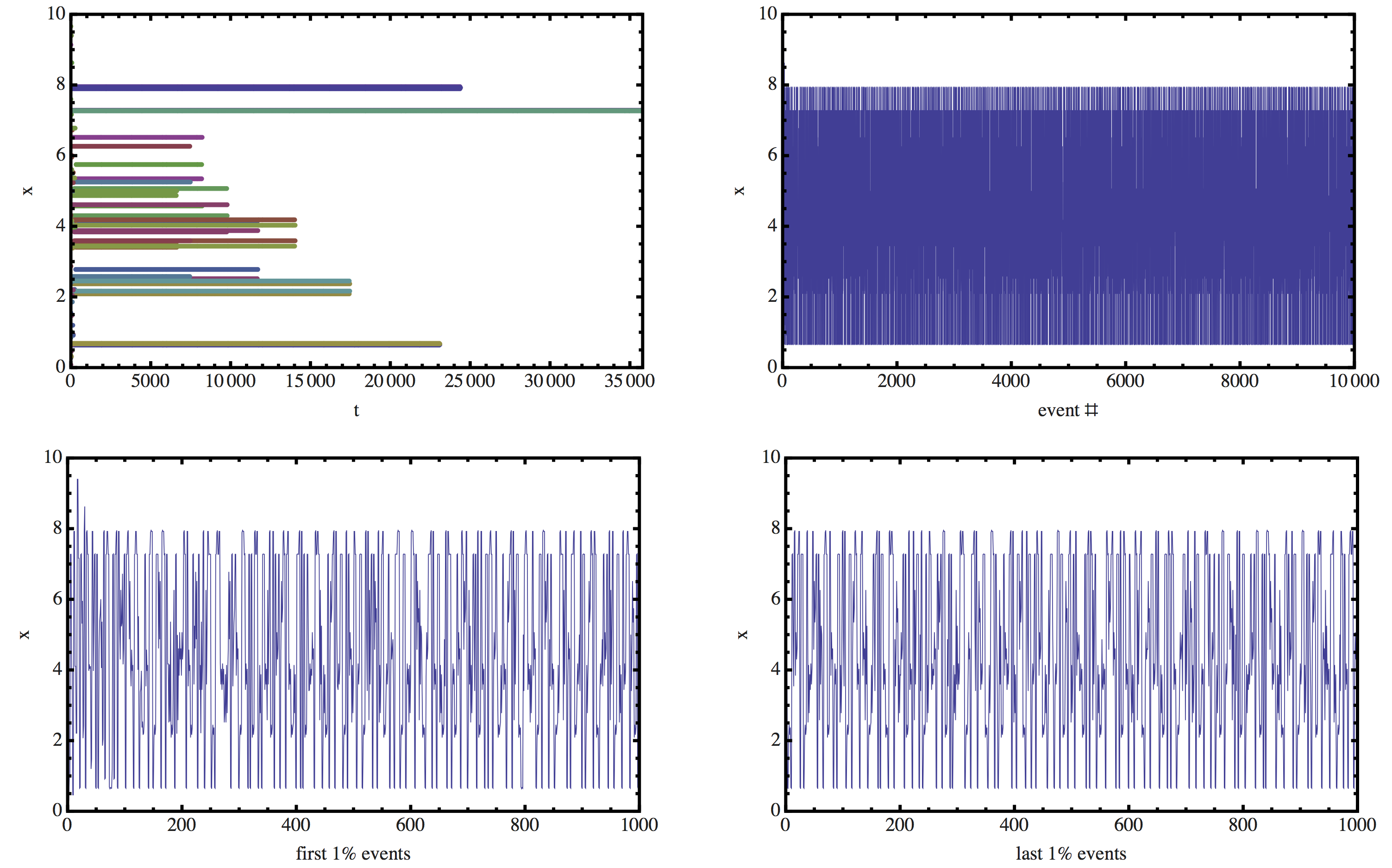}
\caption{Deterministic energetic causal set model, showing the quick capture by a limit cycle. Model with 20 lineages and $10^4$ total steps. In this and the following figures, the story is told in four panels. The upper left panel shows the history plotted in the emergent $1+1$ dimensional spacetime. The upper right panel plots the events in the order by which they take place in the simulation. The lower left panel plots the first 1 percent of events in the sequence and the lower right plots the last 1 percent of events. In these we connect the events in the order generated, which gives us a trajectory.}
\label{DECS}
\end{figure}

Given the $N$ initial events in 
${\cal C}$ there are only $N^2$ possible $x$ positions. 
So the number of possible $z_i^{\pm}$ is finite, and the argument of Section~\ref{finiteDDDS} applies. This means that for the space-time position values $x^a$ the model satisfies all requirements of a finite state deterministic discrete dynamical system and exhibits the convergence to limit cycles. We double check this prediction with simulations of the deterministic version of the ECS model and present the results in Figure~\ref{DECS}.   

We simulate a model with $N=20$ initial events, which means there are 20 different lineages. 
In this and the next figures we describe the simulations results in four panels for each model. The upper left panel plots the coordinates of the emergent space-time position $x$ versus $t$ in ${\cal C}$.
The events are grouped in their lineage by the same color: two events with the same color are in the same lineage. This upper left panel is the same plot we already studied in the 
initial work Ref.~\cite{ECS1}. In the upper right panel we plot the same simulation and events, but plotting the events according to their order in the causal set. This is the order in which they are simulated, step by step as described in the algorithm in section~\ref{algorithm}. In this example the total number of steps is $10^5$. 

The purpose of  studying the event order in the causal set, as opposed to the order in space-time, is so that we can check for repeating patterns in the causal set sequence, and from there look for limit cycles. Limit cycles won't be visible in the space-time diagram. In the lower row we make two zoom-ins of the upper right panel. In the lower panel we plot the first 1\% events in the sequence and in the lower right the last 1\%.

The most stark feature of Figure~\ref{DECS} is the prompt capture of the system by the limit cycles regime. For this deterministic version of the ECS the attraction to the regular phase of limit cycles is immediate. There is a very small number of initial out-of-cycle steps, that we can just distinguish from the bottom left panel displaying the initial 1000 steps. The system then leaves the irreversible phase and collapses promptly to a small number of degrees of freedom, entering the limit cycle regime. Once in the limit cycle the system stays there {\it ad infinitum}. Such behavior holds for simulations of increased number of steps (consistency of the attractor) as well as for increased number of intervenient lineages (increased overall complexity of the causal set). 

We have thus far asserted the convergence to limit cycles of the space-time coordinates $x^a$. This is the first component of the global state $\Psi=\left\{z^{\pm},\Sigma(z^{\pm})\right\}$. To complete the argument we then, equally, have to investigate the capture of the values of the pasts by the limit cycles.  

Since the same considerations of determinism and discreteness apply as those for the positions $x^a$, we are left with the last task of asserting whether the set of possible values for the past $\Sigma$ is finite. 

Like we said above, given the initial $N$ events there are $N^2$ possible positions in ${\cal C}$. The values that $\Sigma(z^{\pm})$ can take are averages of ever increasing numbers of (repeated) elements within this set. Clearly that number is infinite. 

However, in practice, for our system this argument is much simplified, and we show that the number of different values for the pasts is small, and indeed a finite set. Recall that once initiated, the models are promptly captured by limit cycles of typically a small number of elements. This is what we saw in the top left panel of Figure~\ref{DECS}. 

\begin{figure}[h!]
\centering
\includegraphics[width= 0.95\textwidth]{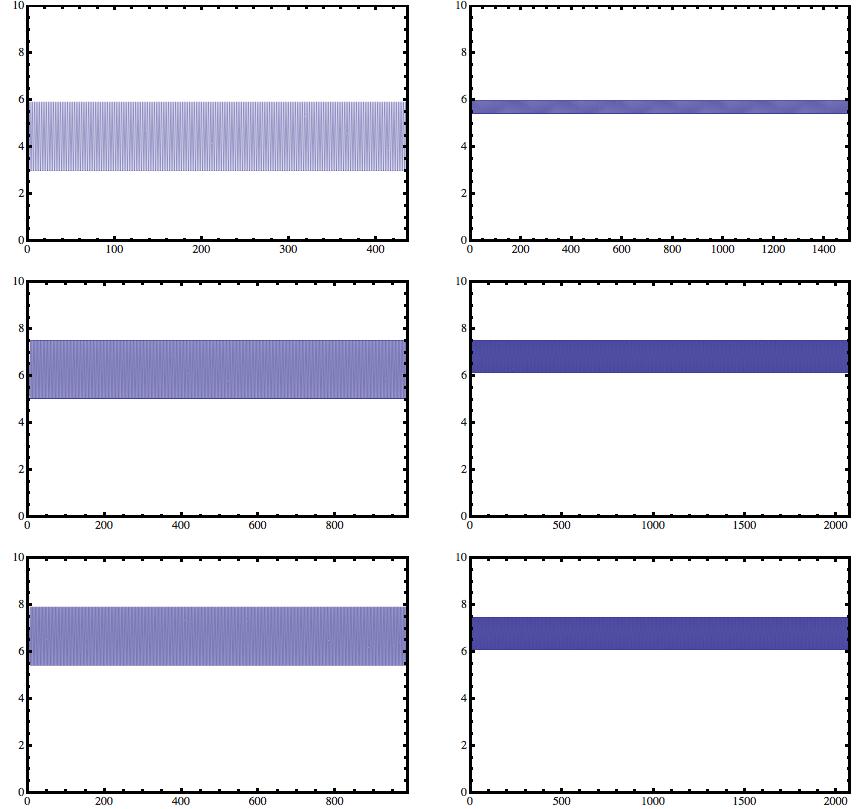}
\caption{Plot of the values of the numerical value of the pasts in each lineage of events, for each step, and for a model with 8 lineages. Note that this is a plot of the value of the pasts of events, and not of their positions $x$, as in other plots. Just like it happens for the value of positions, the evolution of each the pasts also enters a limit cycle promptly. Together with Figure~\ref{DECS} this shows that the pair $\left\{z_i^{\pm}, \Sigma(z_i^{\pm})\right\}$ in the thick present collectively exhibit limit cycles.}
\label{pasts}
\end{figure}

Now we plot the analogous quantity for the values of pasts in each lineage, and at each step in the causal set sequence. In Figure~\ref{pasts} we plot this for a model with 8 lineages, showing each lineage separately. Note that this no longer a plot of the position $x$ but the value of $\Sigma(z^{\pm})$ at each step. 

From Figure~\ref{pasts} we see that 
\begin{enumerate}

\item Initially each lineage jumps around for a handful of steps, before being locked-in a limit cycle of its own; 

\item Each lineage promptly enters the limit cycle regime at the onset of evolutions, and stays there, just like for the space-time positions;

\item This cycle is independent of other families; 

\item For every lineage there are at most only two distinct elements in the limit cycle. 

\end{enumerate}

The last argument strongly limits the number of different values that $\Sigma(z^{\pm})$ can take. This shows that, indeed, the values of the pasts are a finite set, and is consistent with the exhibition of the capture by the limit set. 

Having asserted that both $z^{\pm}$ and $\Sigma(z^{\pm})$ take values in a finest set we can conclude that the number of possible states $\Psi_i$ is finite.


The argument then follows that the deterministic version of ECS exhibits limit cycles. Stronger than this its dynamics is predominantly composed of limit cycles, for a wide range of initial conditions. In particular the space-time positions of the events of the causal set live in the lift of the limit cycles.  They can be visualized as spirals.

Having completed understood the regime of limit cycles for the deterministic model we now proceed the stochastic model behavior. 

\subsection{Converging to limit cycles - stochastic (original) energetic causal set model}
\label{SectionStochECS}

\begin{figure}[t!]
\centering
\includegraphics[width= 0.95\textwidth]{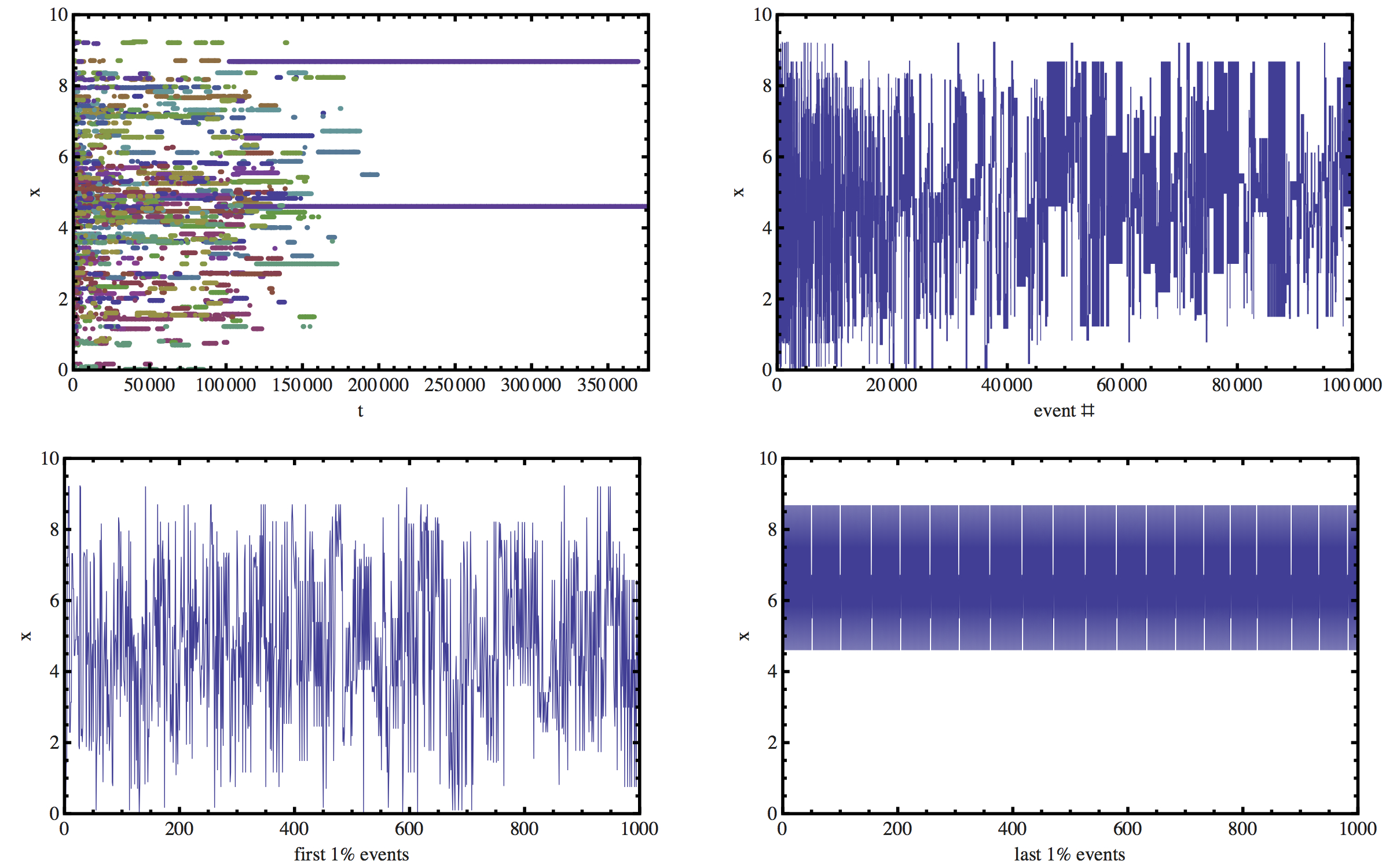}
\caption{
(Number of independent families=20; Number of events $10^5$)
Stochastic energetic causal set model, showing the quick capture by a limit cycle. Panels shown as in Figure~\ref{DECS}. 
In this stochastic model the stark transition between the first 1\% and the last 1\% of events, in this and later figures, depicts clearly the system's evolution towards regularity and limit cycles
}
\label{ECSfig1} 
\end{figure}

We next study the $1+1$ dimensional energetic causal set model we introduced in the 
original work \cite{ECS1}.  Recall from Section~\ref{algorithm} that the dynamics of ECS models is deterministic for the most part, with only a couple of steps where a decision can be made deterministic or random. Those are the choice of the number of windings $n$, and the choice of the pair of momenta to interact. So, even in this original model, the majority of dynamics remains deterministic. However, the small detail of these two stochastic steps has a substantial impact in the complexity of the model and resulting dynamics. 
In Figure~\ref{ECSfig1} we show results from a stochastic ECS model, displaying the same  four panels as in previous figures. 

The effect of the stochastic part of the dynamics, compared to that of determinism in the previous section, can already be seen in the emergent space-time, by comparing the two diagrams in the upper left panels of Figs~\ref{ECSfig1} and \ref{DECS}. The complexity of structure in the stochastic model is significantly higher, there is a initial irregular phase followed by a very regular, where space-time trajectories of quasi-particles emerge. The transition between these two fases is prompt and very clearly marked.
In the initial work in \cite{ECS1}, we observed this transition and two-phase dynamics for a very large range of initial conditions: the emergence of an ordered phase from a disordered phase.  

As mentioned above we now seek to interpret this behaviour observed already in the 
original work in terms of the dynamics of systems which exhibit limit cycles. 

We are now working with a stochastic model so the argument used above for finite state DDDS does not apply. Indeed we would expect that if the system is not deterministic, in general a cycle or repeated pattern will not arise since the uncertainty of the model disturbs it. The argument we gave for the emergence of limit cycles assumes determinism.

Nonetheless, and surprisingly, the stochastic system \textit{does}  converge to limit cycles, contrary to expectation. 
Further, and as we will now explain, a detailed study of the histories of this model shows that the passage from an irregular to a regular phase is explained by the long term behaviour of the model being dominated by limit cycles.

Figure~\ref{ECSfig1} shows an example of a unique events model with $20$ interacting families (or pasts) and $10^5$ total events. In the upper left panel we reproduce the result of the previous work: a plot of physical time on the x-axis by spatial position on the y-axis, with the stark emergence of the quasi-particle trajectories. The remainder panels are as in Figure~\ref{DECS}: they plot the sequence in events of the model in the causal set description. 
The x-axis plots the event number the causal set sequence, that is, the order that they take place in the simulation, and i n the y axis the spatial position of each event. We connect the events in the order generated, which gives us a connected trajectory of jumps at each step, as shown.

\begin{figure}[t!]
\centering
\includegraphics[width= 0.95\textwidth]{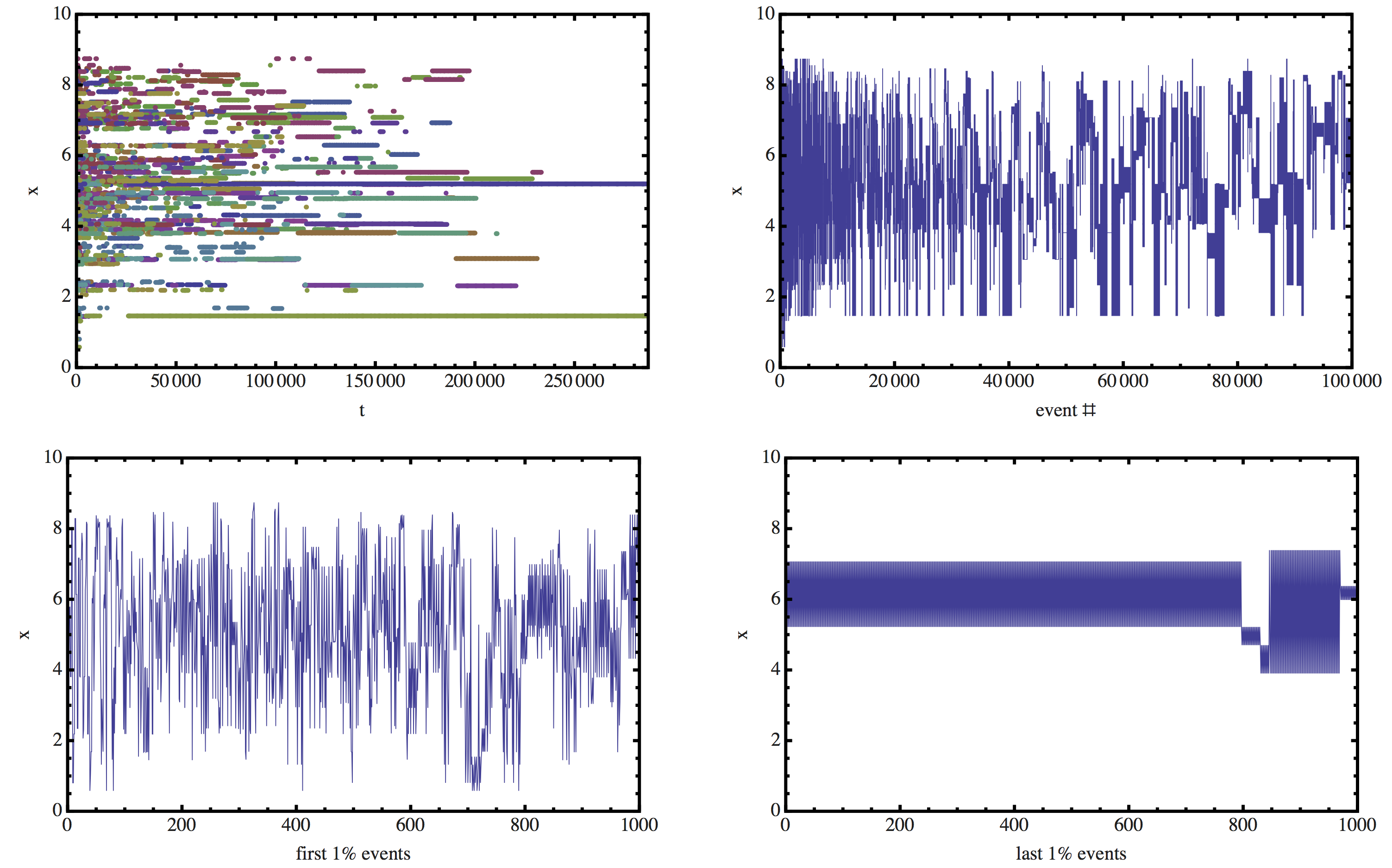}
\caption{ (Number of independent families=20; Number of events $10^5$)
 Here we show an example of another model to illustrate variation between simulation runs.}
\label{ECSfig2} 
\end{figure}

\begin{figure}[h!]
\centering
\includegraphics[width= 0.95\textwidth]{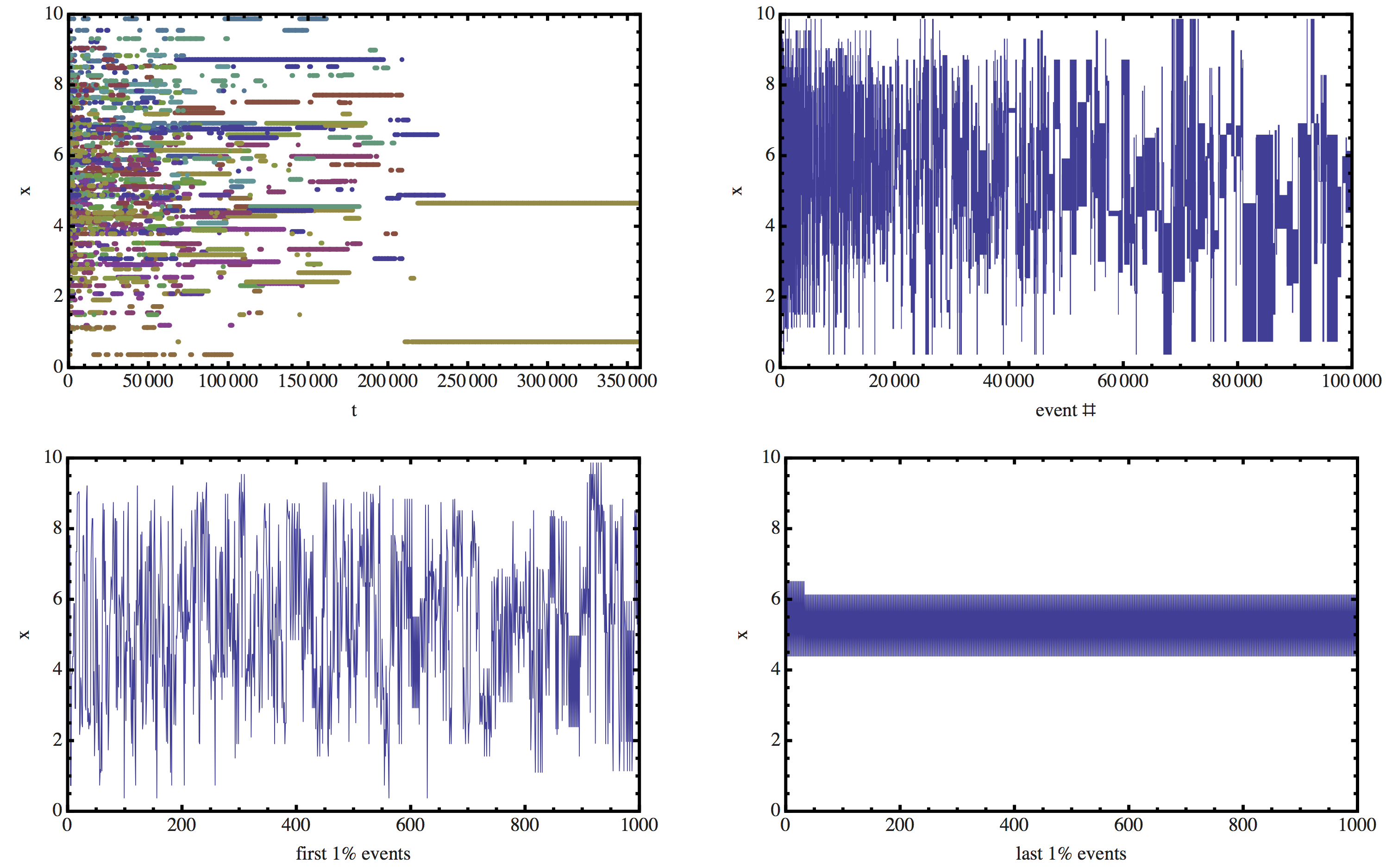}
\caption{ (Number of independent families=20; Number of events $10^5$)
 Here we show an example of another model to illustrate variation between simulation runs.}
\label{ECSfig3} 
\end{figure}

\begin{figure}[t!]
\centering
\includegraphics[width= 0.95\textwidth]{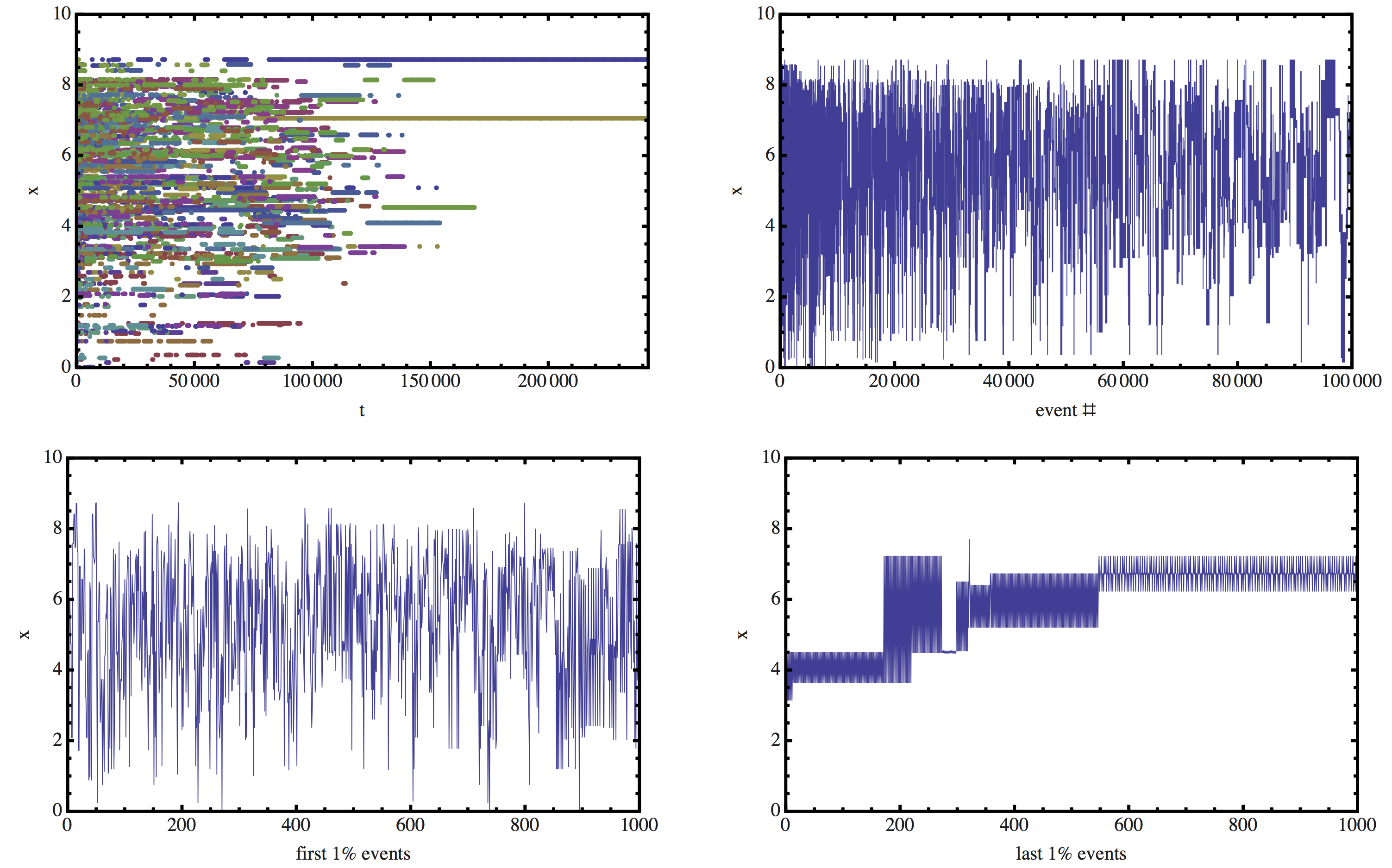}
\caption{(Number of independent families=30; Number of events $10^5$)
If we increase the number of families in the simulation, as expected the time scale for entering the limit cycles is increased.}
\label{ECSfig4} 
\end{figure}

In order to convey variation between runs we give an example in Figures~\ref{ECSfig2} and ~\ref{ECSfig3} of different runs with the same parameters. In Figure~\ref{ECSfig4} we show a simulation with higher number (30) of number of intervening pasts.

Note that, just like for the deterministic case, the order of the events in the causal set description does not necessarily correspond to the emergent space-time order in which they take place, and which is shown in the upper left panel. That is, we don't expect that a plot of order of the events by their spatial position resembles a plot of the physical time vs spatial position (like those in the previous work). In other words, the first row of two panels in each of the figures we present here will be related, but not the same. The first is a plot of events by physical time scale, and the other plots the events in their sequence in the simulation. 

It is  interesting to observe  that a space-time sequence of events need not correspond to the causal set sequence. It is possible that this can be connected to proposals concerning the role of retrocausality in quantum physics\cite{retro}.  This will be addressed in future work 
\cite{AVI+M+L}.

Now we look at the lower pair of panels in the figure.
These are a zoom-in on the sequence of events on the upper right panel. The lower left panel plots the first 1\%  of events in the sequence and the lower right plots the last 1\% of events. This clearly tells the same story as the space-time diagram, the evolution of the system from irregularity towards regularity, this time from another perspective, the order of events in the causal set. 

In Figure~\ref{ECSfig1} (upper right) we observe that initially the events appear to be disordered 
and uncorrelated in spatial position. This is just like the initial phase in the physical time plots (upper left). However, when we plot the actual sequence of events in the simulation (that is, in the DDS language, their trajectory in the space of available states), we observe that after an initial period of jumping around between available states, the sequence gets caught in a repetitive cycle.  It remains trapped in the cycle for number of moves, but not, as would be the  case with a deterministic model, forever.

Instead, in a departure from the usual behaviour of a deterministic DDS, after a certain number of moves the system \textit{exits} the limit cycle and enters a phase of disordered jumping around between states. Further on, it again gets caught in a limit cycle, which may be the same or different than the cycle in which it was initially caught.  Again, after moving cyclically for a while, it jumps out once more.   

This difference in the convergence to the limit cycle between the energetic causal set model and a standard deterministic DDS behaviour arises from the fact that the unique events model is non deterministic.

As it is non deterministic,                                                                                                                                                                                                                                                                                                                                                                                                                                                                                                                                                                                                                                                                                                                                                                                                                                                                                                                                                                                                                                                                                                                                                                                                                                                                                                                                                                                                                                                                                                                                                                                                                                                                                                                                                                                                                                                                                                                                                                                                                                                                                                                                                                                                                                                     we might not have expected that it would exhibit the convergence to limit cycles of the DDS case, because the randomness would be a constant destabilizer of evolution and destabilizer of the cycle. However, contrary to this expectation, we see that this initial disordered evolution between the available states still gives rise to an ordered sequence of states in a limit cycle, because of the dynamics is partly stochastic.

After exiting this initial cycle and going through an amount of disordered evolution for some time, the system again finds another limit cycle in which it will be trapped for some number of steps. In this way the energetic causal set model of \cite{ECS1} goes through a sequence of limit cycles in its evolution, becoming a novel  non-deterministic variation of DDS. 
 
Apart from this, throughout its evolution, the energetic causal set model of \cite{ECS1} evolves progressively from a phase where most of the dynamics is disordered and limit cycles are rare and short (lower left panel), towards a phase where the cycles are longer in duration and the disordered regime between them gets shorter. Towards the end the disordered phase has all but disappeared and the systems jumps between limit cycles (lower right panel). 
The lower panels of each figure illustrate this behaviour, representing the first 1\% (left) and last 1\%(right) of each simulation. 

In order to convey variation between runs we give an example in Figures~\ref{ECSfig2} and ~\ref{ECSfig3} of different runs with the same parameters. In Figure~\ref{ECSfig4} we show a simulation with higher number (30) of number of intervening pasts.

\section{Conclusions}

 
We began by reviewing results, from the study of complex systems, that a large class of irreversible discrete dynamical theories have long term behaviours which are reversible, because their long term behaviour is dominated by limit cycles.  
When the dynamics is restricted to the limit set, the system appears to exhibit time symmetric dynamics, in that the behaviour on the limit set is the same as a reversible dynamics combined with a special initial condition.

We then showed that the dominance of long term behaviour by limit cycles explains the behaviour of the energetic causal set models we introduced \cite{ECS1} which featured the emergence of quasi-particle trajectories.
We have first shown this for the deterministic specialization of ECS, but proceed to obtain the more surprising result that also in their stochastic evolution version the models converge to a limit set.

Thus, the non deterministic energetic causal set model of \cite{ECS1} becomes ``quasi- deterministic" and, hence, also trapped in limit cycles, with the accompanying collapse of the number of degrees of freedom to a finite set.  These systems then show a tendency to evolve towards regularity.  Despite being stochastic, they are attracted to, and get caught in basins of attraction, and spend most of their time in limit cycles.  Because they are stochastic, they occasionally jump out of a limit cycle, but are quickly caught up again in a new one. 

We have thus reinforced the argument of the previous work\cite{ECS1}, that systems which are time irreversible can contain an evolution that is perceived as time reversible. Further, we have shown that those systems can also be interpreted in the framework of discrete dynamical systems. The underlying assertion of both exercises is to show that fundamental equations which are time asymmetric can evolve in such way as to hide the time asymmetry.  

Lastly, and as a result arising from the study of our model through the novel view lens of DDS we arrived at the conclusion that the order of events in the causal set description is not the same as in the emergent space-time, with the overall rule being the preserving of the causal order between the two descriptions. This is work that can perhaps be connected to 
ideas about the role of retrocausality in quantum phenomena\cite{retro}.  This
is currently in progress~\cite{AVI+M+L}.


\section*{Acknowledgements}

We would like to thank Barbara Drossel, Avishalom Elitzur, Andrew Liddle, and Andrew Wuensche for correspondence, encouragement, and comments on the draft. This research was supported in part by Perimeter Institute for Theoretical Physics. Research at Perimeter Institute is supported by the Government of Canada through Industry Canada and by the Province of Ontario through the Ministry of Research and Innovation. M.C.\ was supported by the EU FP7 grant PIIF-GA-2011-300606 and Funda\c{c}\~{a}o para a Ci\^{e}ncia e a Tecnologia (FCT) through grant SFRH/BPD/111010/2015 (Portugal). This research was also partly supported by grants from NSERC and FQXi and we are especially grateful to the John Templeton Foundation for their generous support of this project. Further, this work was also supported by Funda\c{c}\~{a}o para a Ci\^{e}ncia e a Tecnologia (FCT) through the research grant UID/FIS/04434/2013.


\end{document}